\documentclass[letterpaper,twocolumn,10pt]{article}
\usepackage{usenix2019_v3}
\usepackage[compact]{titlesec} 
\usepackage{tikz}
\usepackage{amsmath}
\usepackage{graphicx}
\usepackage{subfig}
\usepackage{hyperref}
\usepackage{float}
\usepackage{enumitem}

\usepackage{filecontents}
\usepackage[sort&compress,numbers]{natbib}

\begin{document}

\date{}

\title{\Large\textbf{Stratum: A Serverless Framework for the Lifecycle
  Management of Machine Learning-based Data Analytics Tasks}}

\author{
{\rm Anirban Bhattacharjee\thanks{These authors contributed equally}, Yogesh Barve\footnotemark[\value{footnote}], Shweta Khare, Shunxing Bao and Aniruddha Gokhale}\\
Vanderbilt University, Nashville, TN, USA
\and
{\rm Thomas Damiano}\\
Lockheed Martin Advanced Technology Laboratories, Cherry Hill, NJ, USA
} 
\let\footnote\thanks

\maketitle

\begin{abstract}
With the proliferation of machine learning (ML) libraries and frameworks, and the programming languages that they use, along with operations of data loading, transformation, preparation and mining, ML model development is becoming a daunting task.   Furthermore, with a plethora of cloud-based ML  model development platforms, heterogeneity in hardware,  increased focus on exploiting edge  computing resources for low-latency prediction serving and often a lack of a complete understanding of resources required to execute ML workflows efficiently, ML model deployment demands expertise for managing the lifecycle of ML workflows efficiently and with minimal cost.  To address these challenges, we propose an end-to-end data analytics, a serverless platform called \emph{Stratum}.  Stratum can deploy, schedule and dynamically manage data ingestion tools, live streaming apps, batch analytics tools, ML-as-a-service (for inference jobs), and visualization tools across the cloud-fog-edge spectrum. This paper describes the Stratum architecture highlighting the problems it resolves.
\end{abstract}

\section{Introduction}

With the increasing availability of data from a variety of sources, and significant improvements in hardware and networks that make Big Data computing easier and affordable, numerous machine learning (ML) libraries and frameworks (e.g., TensorFlow, Scikit Learn, PyTorch) have been designed in the recent past for predictive analytics. Video analysis, Object detection, Speech Recognition, Autonomous cars, Automated traffic signals, industrial robotics are examples of the many real-life applications that demand ML solutions as a part of their live stream analytics or in-depth batch analytics pipeline.   However, writing code for data loading, transformation and pre-processing, and choosing the right ML algorithm for training the data and then evaluating the model and tuning the hyperparameters requires expertise.  The significant promise of using predictive analytics to address a variety of problems of societal and environmental importance~\cite{najafabadi2015deep,batrinca2015social} requires that ML model development be accessible even to novice users. 

Further, there is substantial hype, particularly, with the use of hardware resources (e.g., GPUs, TPUs, FPGAs) along with cloud-offered infrastructure services.  Dealing with this heterogeneity demands expertise in choosing the right hardware configuration that can enhance performance and minimize cost~\cite{venkataraman2016ernest,zhang2017slaq}, which is generally lacking in ML developers.

Consequently, the requirements for lifecycle management of predictive analytics are twofold: 
\setlist{nolistsep}
\begin{enumerate}[itemsep=2pt,leftmargin=15pt]

\item \emph{\textbf{Rapid ML model development framework},} where the goal is to aid ML algorithm developers to build ML models using higher-level abstractions~\cite{hofmann2013rapidminer}.

\item  \emph{\textbf{Rapid ML model deployment framework},} where the goal is to aid developers to deploy and integrate the trained models for analytics on the target hardware and relieve the deployer from having to figure out the right configuration for their ML workflows on the infrastructure~\cite{bhattacharjee2018model}.
	
\end{enumerate}


To that end, we propose a framework called \emph{Stratum}, which addresses the development, deployment, and management lifecycle challenges of data analytics in a heterogeneous distributed environment across the cloud-fog-edge spectrum.  
%
In the rest of this paper, we present the vision behind Stratum, its key features and architectural details in Section~\ref{sec2:arch}, and application areas where Stratum will be useful.

\section{Stratum Vision and Architecture}\label{sec2:arch}

\begin{figure}[htb]
	\vspace{-.1cm}
	\centering
	\includegraphics[width=0.8\linewidth]{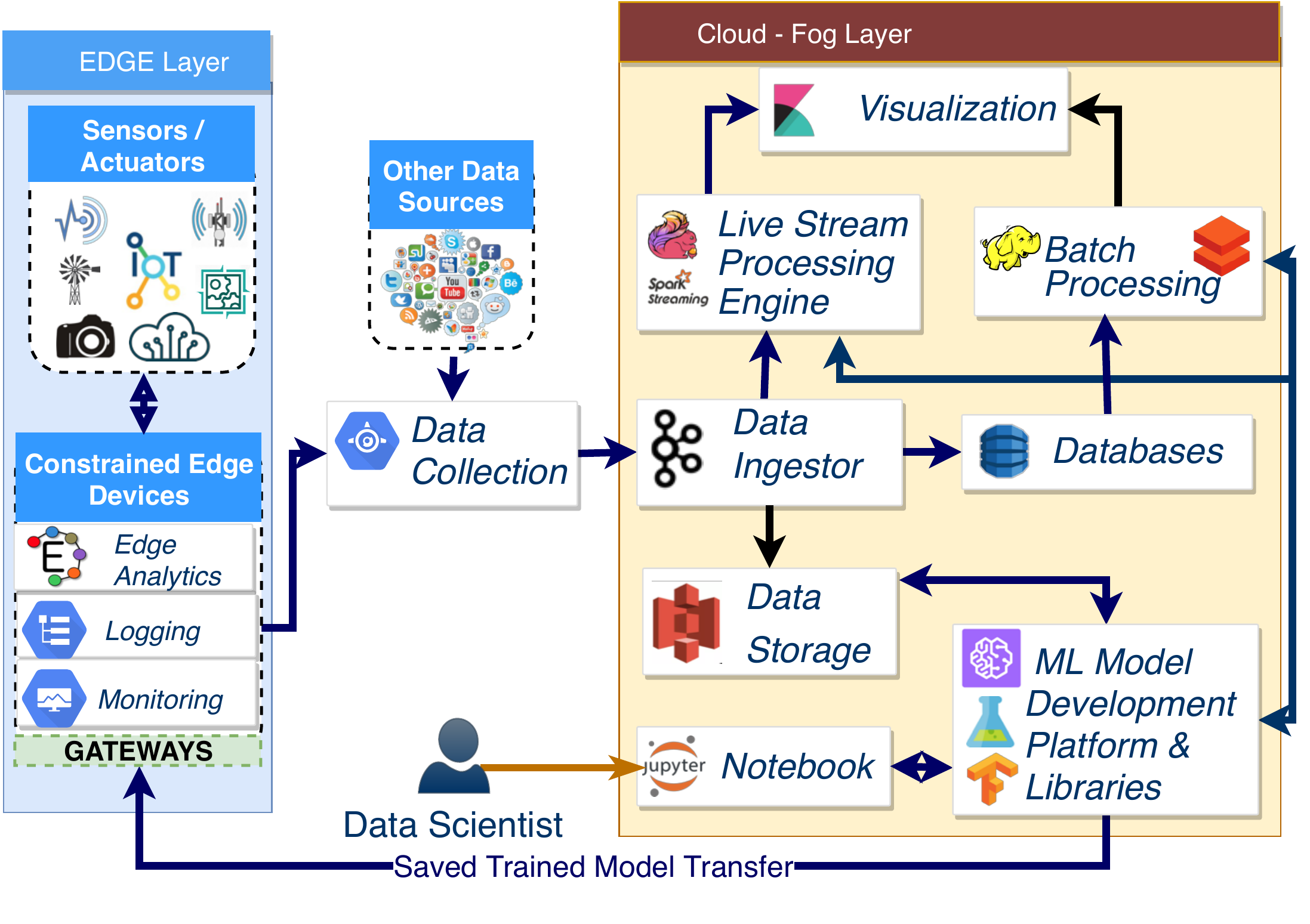}
	\vspace{-.1cm}
	\caption{Generalized Representation of Applications Architecture in Stratum Metamodel}
	\label{Fig:stratum}
	\vspace{-.4cm}
\end{figure}

Figure~\ref{Fig:stratum} depicts the general architecture of how an analytics application can be deployed using Stratum using Model Driven Engineering~\cite{bhattacharjee2018wip}.  We motivate an edge-cloud analytics use case scenario with a smart traffic management system. Traffic cameras collect traffic videos all the time, and rather than sending all the videos to the cloud, edge devices integrated with image recognition capabilities can procure useful insights such as traffic volume, speeding cars and traffic incidents.   Based on data collected over a period of time, the traffic patterns and heavy traffic periods can be learned using batch analytics, which is a computationally intensive process that usually executes in the cloud.  Finally, the intelligent traffic control system typically resides in the fog nodes for real-time needs to dynamically adjust the signal timing of traffic lights based on the learned ML model and by analyzing real-time data using live analytics.   

The Stratum deployment engine can deploy data ingestion tools, stream processing tools, batch analytics tool, machine learning platform, and framework on the target machine (bare metal and virtualized environments) as required. At the heart of Stratum, there is a domain-specific modeling language (DSML) that provides ML developers and deployers a user-interface with higher-level abstractions. 

\begin{figure}[htb]
	\vspace{-.1cm}
	\centering
	\includegraphics[width=0.9\linewidth]{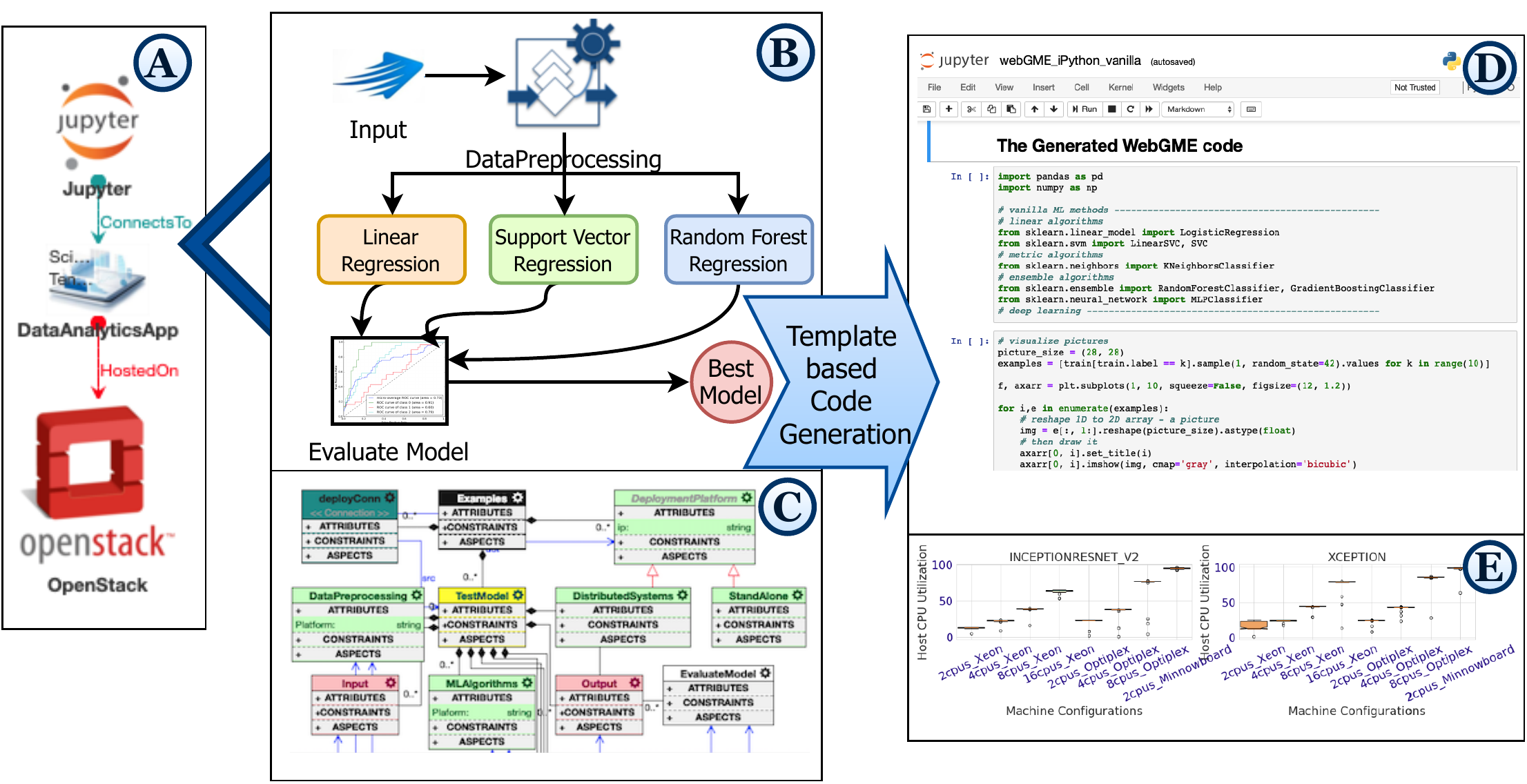}
	\vspace{-.1cm}
	\caption{The user-defined hierarchical model (Blocks A and B) of ML
          model development framework in WebGME), the metamodel (partial) of
          Stratum (Block C), autogeneration of the ML code for subsequent
          deployment and execution (Block D), and performance monitoring tool
          (Block E).}   
	\label{Fig:erudite}
	\vspace{-.4cm}
\end{figure}

Using the DSML, the ML developer can create and evaluate their model using existing ML libraries and frameworks as shown in Figure.~\ref{Fig:erudite}. Based on the user-defined evaluation strategy, Stratum can select the best model by evaluating a series of user-built models. Stratum can distribute each ML model on separate resources to speed up the training and evaluation phase.  Moreover, a Jupyter notebook environment can be attached to our framework so that the auto-generated code by the Stratum DSML can be verified and modified by the expert user if needed.  

Once the ML model is built and evaluated, the Stratum framework can save and profile it.  Stratum supports a pluggable architecture, so the user-supplied specifications are parsed and transformed into deployment-level infrastructure-as-code~\cite{bhattacharjee2018wip,crankshaw2017clipper,gupta2017ifogsim}. Then the user's ML workflows are deployed on the appropriate machines across cloud-fog-edge, and Stratum's serverless execution platform allocates the necessary resources. 
A resource monitoring framework~\cite{barve2018upsara,barve2018poster} within Stratum keeps track of resource utilization and is responsible for triggering actions to elastically scale resources and migrate tasks, as needed, to meet the ML workflow's Quality of Services (QoS). The modeling concepts in Stratum DSML and code generation capabilities of the deployment/management engine are designed using the Web Generic Modeling Environment (WebGME)~\cite{maroti2014next}. Both the DSML and engine are extensible, modularized and reusable. 

\section{Key Features and Benefits of Stratum}

Stratum has been designed with the following key requirements in mind and hence supports the following features:

\setlist{nolistsep}
\begin{enumerate}[itemsep=2pt,leftmargin=15pt]
	
  \item \emph{Rapid Machine Learning (ML) model Development Framework:} The ML model development framework enables fast and flexible deployment of state-of-the-art ML capabilities. It provides a \emph{ML Service Encapsulation} approach leveraging microservice and GPU-enabled containerization architecture and APIs abstracting common ML libraries and frameworks. It provides an easy-to-use scalable framework to build and evaluate ML models. 

  \item \emph{ Rapid Machine Learning (ML) model Deployment Framework: } Stratum provides intuitive and higher-level abstractions to hide the lower-level complexity of infrastructure deployment and management and provides an easy-to-use web-interface for the end users.  The DSML generates ``correct-by-construction''  infrastructure code using constraint checkers before proceeding to actual deployment.  

	\item  \emph{Support for ML Model Transfer: } Stratum provides an intelligent way to transfer the trained model on the target machines (across the cloud-fog-edge spectrum) as an ML module for inference. ML module can be placed on the edge devices,   or it can be placed on Cloud or Fog layer for live or in-depth analysis of data, which depends on user requirements and capacity analysis.  
	
    \item \emph{Extensibility and  Reusability: } Stratum is implemented in a modularized way, and each module is easy to reuse due to plug and play architecture. Similarly, new hardware support can be fused to  Stratum in a standardized manner. 
    
\end{enumerate}


\section*{Availability}
Stratum and its associated tooling are available via Github from 
\url{https://github.com/doc-vu/Stratum}. 

\section*{Acknowledgments}
\small
This work was supported in part by NSF US Ignite CNS 1531079, AFOSR
DDDAS FA9550-18-1-0126 and AFRL/Lockheed Martin StreamlinedML
program. Any opinions, findings, and  conclusions or recommendations
expressed in this material are those of the author(s) and do not
necessarily reflect the views of NSF, AFOSR or AFRL. 

\bibliographystyle{plain}
\bibliography{References}

\begin{thebibliography}{10}

\bibitem{barve2018poster}
Y~Barve, Shashank Shekhar, A~Chhokra, Shweta Khare, Anirban Bhattacharjee, and
  Aniruddha Gokhale.
\newblock Poster: Fecbench: An extensible framework for pinpointing sources of
  performance interference in the cloud-edge resource spectrum.
\newblock In {\em Proceedings of the Third ACM/IEEE Symposium on Edge
  Computing}, 2018.

\bibitem{barve2018upsara}
Yogesh Barve, Shashank Shekhar, Shweta Khare, Anirban Bhattacharjee, and
  Aniruddha Gokhale.
\newblock Upsara: A model-driven approach for performance analysis of
  cloud-hosted applications.
\newblock In {\em 2018 IEEE/ACM 11th International Conference on Utility and
  Cloud Computing (UCC)}, pages 1--10. IEEE, 2018.

\bibitem{batrinca2015social}
Bogdan Batrinca and Philip~C Treleaven.
\newblock Social media analytics: a survey of techniques, tools and platforms.
\newblock {\em Ai \& Society}, 30(1):89--116, 2015.

\bibitem{bhattacharjee2018model}
Anirban Bhattacharjee, Yogesh Barve, Aniruddha Gokhale, and Takayuki Kuroda.
\newblock A model-driven approach to automate the deployment and management of
  cloud services.
\newblock In {\em 2018 IEEE/ACM International Conference on Utility and Cloud
  Computing Companion (UCC Companion)}, pages 109--114. IEEE, 2018.

\bibitem{bhattacharjee2018wip}
Anirban Bhattacharjee, Yogesh Barve, Aniruddha Gokhale, and Takayuki Kuroda.
\newblock (wip) cloudcamp: Automating the deployment and management of cloud
  services.
\newblock In {\em 2018 IEEE International Conference on Services Computing
  (SCC)}, pages 237--240. IEEE, 2018.

\bibitem{crankshaw2017clipper}
Daniel Crankshaw, Xin Wang, Guilio Zhou, Michael~J Franklin, Joseph~E Gonzalez,
  and Ion Stoica.
\newblock Clipper: A low-latency online prediction serving system.
\newblock In {\em NSDI}, pages 613--627, 2017.

\bibitem{gupta2017ifogsim}
Harshit Gupta, Amir Vahid~Dastjerdi, Soumya~K Ghosh, and Rajkumar Buyya.
\newblock ifogsim: A toolkit for modeling and simulation of resource management
  techniques in the internet of things, edge and fog computing environments.
\newblock {\em Software: Practice and Experience}, 47(9):1275--1296, 2017.

\bibitem{hofmann2013rapidminer}
Markus Hofmann and Ralf Klinkenberg.
\newblock {\em RapidMiner: Data mining use cases and business analytics
  applications}.
\newblock CRC Press, 2013.

\bibitem{maroti2014next}
Mikl{\'o}s Mar{\'o}ti, Tam{\'a}s Kecsk{\'e}s, R{\'o}bert Keresk{\'e}nyi, Brian
  Broll, P{\'e}ter V{\"o}lgyesi, L{\'a}szl{\'o} Jur{\'a}cz, Tihamer
  Levendovszky, and {\'A}kos L{\'e}deczi.
\newblock Next generation (meta) modeling: Web-and cloud-based collaborative
  tool infrastructure.
\newblock {\em MPM@ MoDELS}, 1237:41--60, 2014.

\bibitem{najafabadi2015deep}
Maryam~M Najafabadi, Flavio Villanustre, Taghi~M Khoshgoftaar, Naeem Seliya,
  Randall Wald, and Edin Muharemagic.
\newblock Deep learning applications and challenges in big data analytics.
\newblock {\em Journal of Big Data}, 2(1):1, 2015.

\bibitem{venkataraman2016ernest}
Shivaram Venkataraman, Zongheng Yang, Michael~J Franklin, Benjamin Recht, and
  Ion Stoica.
\newblock Ernest: Efficient performance prediction for large-scale advanced
  analytics.
\newblock In {\em NSDI}, pages 363--378, 2016.

\bibitem{zhang2017slaq}
Haoyu Zhang, Logan Stafman, Andrew Or, and Michael~J Freedman.
\newblock Slaq: quality-driven scheduling for distributed machine learning.
\newblock In {\em Proceedings of the 2017 Symposium on Cloud Computing}, pages
  390--404. ACM, 2017.

\end{thebibliography}

\end{document}